\documentstyle[epsfig,twocolumn,fleqn]{article}

\begin{document}
\renewcommand{\textfraction}{0.1}
\renewcommand{\topfraction}{0.8}
\rule[-8mm]{0mm}{8mm}
\begin{minipage}[t]{16cm}
\begin{center}
{\Large \bf Local mode behaviour in quasi-1D CDW systems \\[4mm]}
H. Fehske$^{\rm a}$, G.~Wellein$^{\rm b}$, H.~B\"uttner$^{\rm a}$, 
A. R. Bishop$^{\rm c}$, and M. I. Salkola$^{\rm d}$\\[3mm]
$^{\rm a}${Physikalisches Institut, Universit\"at Bayreuth, 
D-95440 Bayreuth, Germany}\\
$^{\rm b}${Regionales Rechenzentrum Erlangen, Universit\"a{}t Erlangen, 
  91058 Erlangen, Germany }\\
$^{\rm c}${MSB262, Los Alamos National Laboratory, Los Alamos,
  NM 87545, USA}\\
$^{\rm d}${Superconductor Technologies Inc., Santa Barbara, CA 93111, USA}
\\[4.5mm]
\end{center}
{\bf Abstract}\\[0.2cm]
\hspace*{0.5cm}
We analyze numerically the ground-state and spectral properties of 
the three-quarter filled Peierls-Hubbard Hamiltonian. 
Various charge- and spin-ordered states are identified.
In the strong-coupling regime, we find clear signatures 
of local lattice distortions accompanied by intrinsic 
multi-phonon localization. The results are discussed 
in relation to recent experiments on MX chain  [-PtCl-] complexes.
In particular we are able to reproduce the observed red shift of 
the overtone resonance Raman spectrum.\\[0.2cm] 
  \end{minipage}\\[4.5mm]
\normalsize
Inspired by the recent observation of intrinsically localized vibrational
modes in halide-bridged transition metal [-PtCl-] complexes~[1], 
we study strong coupling effects between electronic and lattice degrees of 
freedom on the basis of a two-band, 3/4-filled  Peierls-Hubbard 
model (PHM)
\begin{eqnarray}
{\cal H}\!\!\!&=&\!\!\!
-t\sum_{\langle i,j \rangle \sigma} c_{i\sigma}^{\dagger}c_{j\sigma}^{}
+\sum_{i\sigma} \varepsilon_i n_{i\sigma}^{}+ \sum_{i} U_i n_{i\uparrow}^{}n_{i\downarrow}^{}\nonumber\\
&&\hspace*{-0.5cm}
+\lambda_{I}\, (b_{I}^{}+b_{I}^{\dagger})(n_1+n_3-n_2-n_4)
+\hbar\omega_{I}\, b_{I}^{\dagger}b_{I}^{}
\nonumber\\&&\hspace*{-0.5cm}
+\lambda_R \,(b_{R}^{}+b_{R}^{\dagger})(n_2-n_4)
+\hbar\omega_{R}\, b_{R}^{\dagger}b_{R}^{}\,.
\end{eqnarray} 
In~(1),  the $c_{i\sigma}^{[\dagger]}$ are fermion operators,
$n_{i\sigma}=c_{i\sigma}^{\dagger}c_{i\sigma}^{}$, and  $b_{R}^{[\dagger]}$
and $b_{I}^{[\dagger]}$ are the boson operators for the 
Raman active (R)  and  infrared (I) optical phonon modes
with bare phonon frequencies $\omega_{R}$ and  $\omega_{I}$.
With regard to the quasi-1D charge density wave (CDW) system
$\rm \{[Pt(en)_2] [Pt(en)_2Cl_2](ClO_4)_4\}$ (en= ethylenediamine),
subsequently abbreviated as PtCl, the Pt (Cl) atoms are denoted by the 
site index $i=2,\,4$ ($i=1,\,3$). 
The CDW state is built up by alternating nominal
$\rm Pt^{+4}$ and $\rm Pt^{+2}$ sites with a corresponding distortion of 
Cl$^-$-ions towards $\rm Pt^{+4}$. To model the situation of a  3/4-filled
charge transfer insulator  within a {\it single-mode approach} 
(SMA), we only include the R-($\nu_1$)-mode 
with $\hbar\omega_R=0.05$~[2], and parametrize 
the site energies by 
$\Delta=(\varepsilon_{\rm Pt}-\varepsilon_{\rm Cl})/t=1.2$ and 
the Hubbard repulsions by $U_{\rm Pt}=0.8$ and $U_{\rm Cl}=0$ 
(all energies are measured in units of $t$).   
Alternatively, we employ a more realistic {\it double-mode} 
{\it approach} (DMA), using $\hbar\omega_I=0.06$ for the 
I-($\nu_2$)-mode~[2].
\begin{figure}[t]
\unitlength1mm
\begin{picture}(70,75)
\end{picture}
\end{figure}
\begin{figure}[h]
\centerline{\mbox{\epsfig{file= 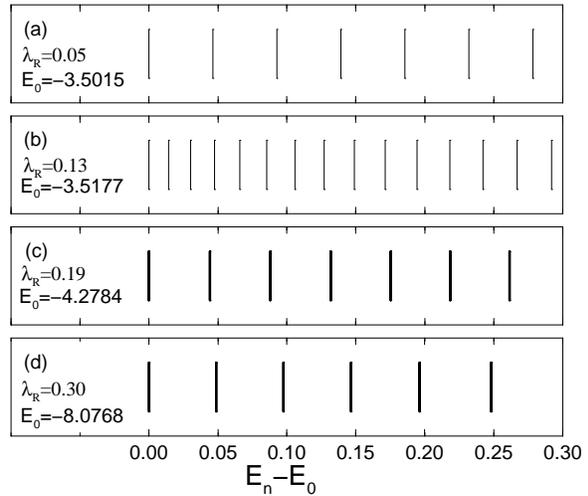, width =\linewidth}}}\vspace*{-0.6cm}
\caption{\small Low-energy part of the eigenvalue spectrum of the  
PHM [periodic boundary conditions, SMA]. 
Eigenvalues in (c) and (d) are two-fold degenerate.}
\end{figure}

The ground-state and spectral properties of the four-site 
PHM are determined by 
finite-lattice Lanczos diagonalization that preserves 
the full dynamics of the phonons~[3].

Figure~1 shows the variation of the lowest energy states 
as a function of the electron-phonon  coupling.
In the weak-coupling region, the ground-state is 
basically a zero-phonon state (cf.
\begin{table}[t]\vspace*{-0.4cm}
\caption{Relative red shift of the lowest-energy peaks
in dependence on the final quanta of the vibrational 
energy [$\lambda_R=0.19$; SMA].}
\vspace*{0.2cm}
\begin{tabular}{lrrrrrr} \hline\hline
\multicolumn{1}{c}{$r_n$ [\%]\rule{0mm}{4mm} }&\multicolumn{1}{c}{$n=2$} &\multicolumn{1}{c}{ 3} &\multicolumn{1}{c}{ 4 }&\multicolumn{1}{c}{ 5} &\multicolumn{1}{c}{ 5} &\multicolumn{1}{c}{ 7}\\ \hline\hline
Pt$^{37}$Cl~[1]\rule{0mm}{4mm}  &  0.4 &  1.1 &  2.4 &  4.6 & 7.7 & 11.6\\
SMA \rule{0mm}{4mm} &  0.4 &  1.3 & 2.8 & 4.8 & 7.5 & 10.9\\ 
\hline\hline
\end{tabular}
\caption{\mbox{Different energy contributions to 
$E_0$,} 
local particle densities [$\langle n_i\rangle$],
local magnetic moments [$L_i= 3\langle (n_{i\uparrow}
-n_{i\downarrow})^2\rangle/4 $], charge correlations 
[$\langle n_i n_j\rangle$] and magnetic structure factor 
[$S_s(\pi)=\frac{1}{N}\sum_{i,j}\langle S_i^z S_j^z
\rangle \mbox{e}^{i\pi (i-j)}$] in the ground-state
of the PHM. Note that the effect of $\Delta$  
can be obtained dynamically within the DMA (see column 4).}
\vspace*{0.2cm}
\begin{tabular}{lrrrr}\hline\hline
\multicolumn{1}{l}{ \rule{0mm}{4mm}}&
\multicolumn{2}{c}{SMA} &
\multicolumn{1}{c}{DMA} &
\multicolumn{1}{c}{$\Delta=0$}   \\
\multicolumn{1}{l}{$\lambda_{I}$ \rule{0mm}{4mm}} & \multicolumn{1}{c}{0.0}  
&  \multicolumn{1}{c}{ 0.0} & \multicolumn{1}{c}{0.15} 
& \multicolumn{1}{c}{ 0.15} \\
\multicolumn{1}{l}{$\lambda_{R}$ \rule{0mm}{4mm}} & \multicolumn{1}{c}{0.05}  
&  \multicolumn{1}{c}{ 0.19} & \multicolumn{1}{c}{ 0.19} 
& \multicolumn{1}{c}{ 0.05} 
\\ \hline\hline 
$E_0$ \rule{0mm}{4mm}         &   -3.5015    &  -4.2783  & -5.3770 & -3.2715 \\
$E_{kin}$      &   -4.1115  & -2.9186  &  -2.3202 &  -3.5985 \\
$E_{ph} $      &   0.0001     &  2.1485   & 3.8490 & 0.2602  \\
$E_{el-ph} $   &  -0.0036     & -4.3026  & -7.7017 & -0.5433  \\
$E_{U}$        &   0.6136   &  0.7945   & 0.7959 &  0.6102  \\[0.1cm]
\hline
$\langle n_{\rm Cl}\rangle $& 1.7047 & \rule{0mm}{4mm} 1.8834  &  1.9561
&  1.7070 \\
$\langle n_{\rm Pt(2)}\rangle$  &  1.2953 & 0.2541 & 0.0949 &1.2929  \\
$\langle n_{\rm Pt(4)}\rangle$  &  1.2953 & 1.9791  & 1.9929 &1.2929 \\[0.1cm]
\hline
$\langle n_1 n_2\rangle$ \rule{0mm}{4mm} & 2.1053  & 0.3788 & 0.1441 &2.1042 \\
$\langle n_1 n_3\rangle$  &  2.8611 &   3.5396 & 3.8253 &2.8705   \\
$\langle n_2 n_4\rangle$  &  1.4990  &  0.4852 & 0.1824 &1.4937   \\[0.1cm]
\hline
$L_{\rm Cl}$\rule{0mm}{4mm}  & 0.1897 &  0.0830 &  0.0323  & 0.1880\\
$L_{\rm Pt(2)} $      &  0.3962    &  0.1696 &  0.0682  & 0.3977 \\
$L_{\rm Pt(4)} $      &  0.3962    &  0.0156 &  0.0053  &0.3977  \\[0.1cm]
\hline 
$S_s(\pi)$   \rule{0mm}{4mm}      & 0.1053 
  &  0.0524  & 0.0212 & 0.1042 \\[0.1cm]
\hline
Config.:\rule{0mm}{4mm}  & $\bullet$-$\uparrow$-$\bullet$-$\downarrow$ & 
$\bullet$-$\circ$-$\bullet$-$\bullet$ &$\bullet$-$\circ$-$\bullet$-$\bullet$& 
$\bullet$-$\uparrow$-$\bullet$-$\downarrow$
\\[0.1cm]
\hline\hline
\end{tabular}
\end{table}
Fig.~2~a) and the peaks  
in Fig.~1~(a) correspond to multiples of the fundamental
phonon frequency $\omega_R^{(1)}$. As $\lambda_R$ increases 
a strong mixing of electron and phonon degrees of freedom takes place, 
and finally the lowest states with total momentum ${\bf} K=0$ and  
${\bf} K=\pi$ become nearly degenerate. 
In this limit a non-linear lattice potential stabilizing 
so-called {\it intrinsically localized vibrational modes}   
(ILMs) is self-consistently generated~[1].
\begin{figure}[t]
\centerline{\mbox{\epsfig{file= 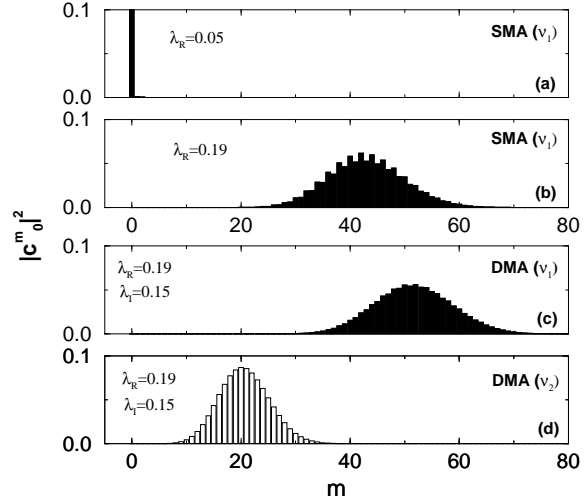, width =\linewidth}}}\vspace*{-0.6cm}
\caption{\small Phonon distribution in the ground state   
for the antisymmetric I (white columns) and symmetric Cl-Pt-Cl stretch
R (black columns) phonon modes.}
\end{figure}
A prominent feature of these bound states is their strong
anharmonic redshift, resulting from the attractive interaction 
of Raman phonon quanta located at the same PtCl$_2$ unit.
The calculated red shift, 
$r_n=[n \omega_R^{(1)} - \omega_R^{(n)}]/\omega_R^{(1)}$ with 
$\omega_R^{(n)}= (E_n-E_0)$, of the (doublet) overtones shown in Fig.~1~(c)
is successfully compared to experimental data probed by 
resonance Raman scattering (see Tab.~1).
To elucidate the different nature of the ground state in the weak- and
strong-coupling regimes, several characteristic quantities 
listed in Tab.~2. Obviously the self-localization transition 
is accompanied by significant changes in the spin- and charge correlations.
As can be seen from the weight of the $m$-phonon state in the ground state,
$|c_0^m|^2$~[3], depicted in Fig.~2, the appearance of the  
$\bullet$-$\circ$-$\bullet$-$\bullet$ configuration 
is related to large occupation numbers of the localized 
vibrational (R) mode. 

To summarize, the numerical results obtained for the PHM 
provide strong evidence for the existence of a dynamical 
spatial localization of vibrational energy (ILM) in MX solids,
due to high intrinsic nonlinearity 
from strong electron lattice coupling.
{\small 
\begin{enumerate}
\item[{[1]}]
B. I. Swanson et al., Phys. Rev. Lett. {\bf 82} (1999) 3288.
\item[{[2]}]
S. P. Love  et al., Phys. Rev. B {\bf 47} (1993) 11107.
\item[{[3]}]
B. B\"auml et al., Phys. Rev. B {\bf 98} (1998) 3663.
\end{enumerate}
}
\end{document}